\documentclass[epsf]{aa}
\usepackage{graphicx}

\begin{document}
\def\fu{$f_1$}
\def\t{$\pm$}
\def\ft{$f_3$}
\def\fq{$f_4$}
\def\fdu{$f_2 - 2f_1$}
\def\fp{$f_1 + f_2$}
\def\fm{$f_2 - f_1$}
\def\cd{d$^{-1}$}
\def\cds{d$^{-1}$\,}
\def\kms{km~s$^{-1}$}
\def\gsc{GSC~00144-03031 }
\def\gscp{GSC~00144-03031}
\def\kmss{km~s$^{-1}$\,}
\def\I{\'\i}
\def\salp{\vskip 0.3truecm}
\title{The double-mode nature of the HADS star GSC~00144-03031 and
the Petersen diagram of the class
\thanks{Based on observations partly collected
at S.~Pedro M\'{a}rtir and Sierra Nevada Observatories}}
\author{E.~Poretti\inst{1}, J.C.~Su\'{a}rez\inst{2}, P.G.~Niarchos\inst{3}, K.D.~Gazeas\inst{3}, 
V.N.~Manimanis\inst{3}, P.~Van Cauteren\inst{4,5}, P.~Lampens\inst{6}, P.~Wils\inst{7}, 
R.~Alonso\inst{8}, P.J.~Amado\inst{2,9}, J.A.~Belmonte\inst{8}, N.D.~Butterworth\inst{10}, 
M.~Martignoni\inst{5}, S.~Mart\I n-Ruiz\inst{2},
P.~Moskalik\inst{11}, and C.W.~Robertson\inst{12}}
\institute {
INAF-Osservatorio Astronomico di Brera, Via Bianchi 46,
I-23807 Merate, Italy\\
 \email{poretti@merate.mi.astro.it}
\and Instituto de Astrof\I sica de Andaluc\I a, C.S.I.C., Apdo. 3004, 18080 Granada,
Spain
\and Department of Astrophysics, Astronomy and Mechanics, Faculty of Physics, University of Athens, Panepistimiopolis,
15784 Zografos, Athens, Greece
\and Beersel Hills Observatory, Laarheidestraat 166, 1650 Beersel, Belgium
\and Groupe Europ\'{e}en d'Observations Stellaires (GEOS), 23 Parc de Levesville, F-28300 Bailleau l'Ev\"{e}que, France
\and Koninklijke Sterrenwacht van Belgi\"{e}, Ringlaan 3, 1180 Brussel, Belgium
\and Vereniging Voor Sterrenkunde, Belgium
\and Instituto de Astrof\I sica de Canarias, C/ V\I a L\'actea s/n, 38200 La
Laguna, Tenerife, Spain
\and European Southern Observatory, Alonso de Cordova 3107, Santiago 19, Chile
\and 24 Payne Street, Mt. Louisa, Townsville, Queensland 4814, Australia
\and Nicolaus Copernicus Astronomical Center, ul. Bartycka 18, 00-716 Warsaw, Poland
\and SETEC Observatory, Goddard, KS, USA
}
\offprints{E. Poretti}
\date{Received date; Accepted Date}
\abstract{The double--mode pulsation of \gsc has been detected when searching for COROT
targets. A very large dataset composed of 4722 photometric measurements was  collected
at six observatories in Europe and America. There is no hint of the excitation of 
additional modes (down to 0.6 mmag) and therefore \gsc seems to be a pure double--mode pulsator,
with a very short fundamental radial mode ($P$=84~min).
From $uvby\beta$ photometry and evolutionary tracks it appears to be a Pop.~I star
with M=1.75~M$_{\sun}$, located in the middle of the instability strip, close to the Zero--Age
Main Sequence.  
We also discovered other new double--mode pulsators in the databases of large--scale
projects: OGLE BW2\_V142, OGLE BW1\_V207, ASAS3 094303-1707.3, ASAS3 000116-6037.0,
NSVS 3234596 and NSVS 3324715.  An observational Petersen diagram is presented and 
explained by means of new models. 
A common sequence connecting Pop.~I stars from the shortest to the longest periods is proposed and  
the spreads in the period ratios are ascribed to different metallicities (at the shortest periods)
and to different masses (at the longest ones).
}
\authorrunning{E. Poretti et al.} 
\titlerunning{GSC 00144-03031 and double--mode HADS stars}
\maketitle

\section{Introduction} 

The double--mode High Amplitude Delta Scuti (HADS) stars have been considered
for a long time 
similar to double--mode Cepheids, i.e., evolved stars pulsating
in the first two radial modes only. A remarkable 
exception is constituted
by the AC And subclass, composed of stars  pulsating in the first three
radial modes (see Tab.~9 in Rodr\I guez 2003). Very recently, nonradial 
modes have been found combined with
the radial ones, making these stars more similar to the low--amplitude,
unevolved $\delta$~Sct stars which are quite common between the
Zero--Age and the Terminal--Age Main Sequences (Poretti 2003). Therefore,
HADS stars lost their definition of ``simple double-- or single--mode radial 
pulsators". A natural
question arises:  do all the double--mode HADS stars display nonradial
modes? To give a first answer to this question we investigated
in detail the light curve of \gscp,  a new member of the class. Moreover,
since other double--mode pulsators have been found in large--scale
projects, it is timely    
to discuss the entire class by comparing the new, more robust  observational results
with the theoretical  models.

The variability of \gsc was discovered when observing the field around
HD~43587, a candidate primary target for the asteroseismic space 
mission COROT (COnvection, ROtation and planetary Transits; Baglin et al. 2002). 
The field around this star has been explored in the search for secondary
targets (Poretti et al. 2005) and the STARE telescope 
\footnote{http://www.hao.ucar.edu/public/research/stare/stare.html}
detected the variability
of \gsc on the night of 16--17 March 2002 (JD 2452350). The STARE light curve
clearly showed the effect of multiperiodicity in this large amplitude variable
($\Delta V >$0.40 mag).

\section{Observations}

\begin{figure}
\resizebox{\hsize}{!}{\includegraphics{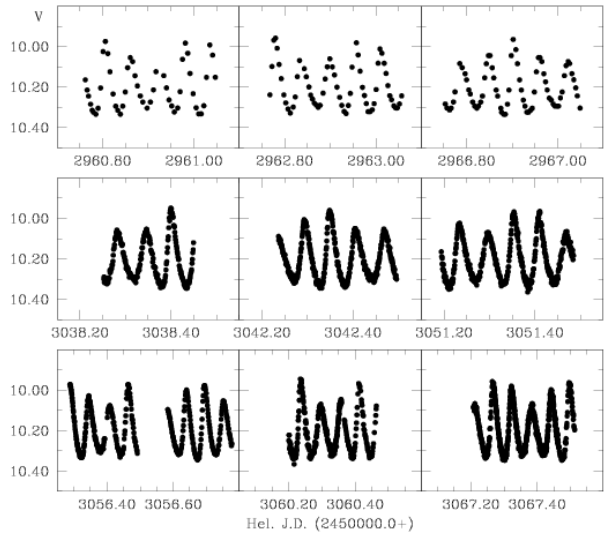}}
\caption[ ]{Light curves of \gsc  in standard $V$ magnitudes. The panels
in the upper row show photoelectric data obtained at S.~Pedro M\'{a}rtir
Observatory. The panels in the middle row show CCD data obtained at
Athens University Observatory. The panels in the lower row are, from left:
CCD data obtained at Beersel Hills and SETEC Observatories
(JD. 2453056); 
CCD data obtained at Athens University Observatory  and by MM (JD 2453060);
CCD data obtained at Beersel Hills and  at Athens University 
Observatories (JD 2453067).
}
\label{lc}
\end{figure}

Owing to the recent growing interest in high--amplitude  pulsators, new 
observations were planned to investigate the pulsational content
of \gsc in detail.
The first long subset was obtained at
S.~Pedro M\'{a}rtir Observatory: 249  measurements were collected using the
simultaneous $uvby$ Danish photometer mounted on the 1.5--m 
telescope, on five nights from November 14 to 23, 2003. The observations were
performed with the scope of monitoring the brightest COROT targets and
therefore the time sampling is not very dense (Fig.~\ref{lc}, upper row).
HD~43913 was used as comparison star. Later in the same winter, dedicated intensive 
CCD monitoring was 
performed at Athens University Observatory: a 0.40--m telescope equipped with a 
SBIG ST-8 camera and a $V$ filter was used on eight nights from January 26, 2004
to March 2, 2004. GSC 00144-02970 was  used as comparison star and
GSC 00144-03140 as check star: both are located in the same field of view
and the monitoring was
practically continuous (Fig.~\ref{lc}, middle row). A third 
dataset was   obtained at Beersel Hills
Observatory (Belgium), using a 0.40--m telescope equipped with an ST10
XME CCD camera and a $V$ filter. The observations were performed on ten nights
from  February 16, 2004 to March 28, 2004 (Fig.~\ref{lc}, lower row). 
Other less conspicuous subsets were collected at SETEC Observatory
(0.30--m telescope equipped with a ST--8 CCD camera and a $V$ filter, 
168 measurements on one night), at
Sierra Nevada Observatory  (OSN; 116 $uvby$ data obtained at the 0.90--m 
telescope on three nights) and by MM (0.20--m telescope
equipped with  a Kodak 401 CCD camera and a $V$ filter, 483 measurements on four nights).
The time baseline spans 133~d.

\section{Frequency analysis}

\begin{table*}
\caption{Parameters of the least--squares fits of the measurements on
\gscp. 
$T_0$=HJD~2453054.1189}
\begin{tabular} {l c r  c  rr c rr c rr  c rr }
\hline
\hline
\noalign{\smallskip}
& & & & \multicolumn{2}{c}{Global dataset}& & \multicolumn{2}{c}{SPM subset} & &
\multicolumn{2}{c}{BHO subset}& & \multicolumn{2}{c}{AUO subset}\\
\cline{5-6}\cline{8-9}\cline{11-12}\cline{14-15}
\multicolumn{1}{c}{Term} & & \multicolumn{1}{c}{Freq.} & &\multicolumn{1}{c}{Ampl.} &
\multicolumn{1}{c}{Phase} & &\multicolumn{1}{c}{Ampl.} & \multicolumn{1}{c}{Phase} & &\multicolumn{1}{c}{Ampl.} &
\multicolumn{1}{c}{Phase}& & \multicolumn{1}{c}{Ampl.} & \multicolumn{1}{c}{Phase} \\
\multicolumn{1}{c}{} & & \multicolumn{1}{c}{[\cd]} & & \multicolumn{1}{c}{[mag]} & \multicolumn{1}{c}{[rad]} & &
\multicolumn{1}{c}{[mag]} & \multicolumn{1}{c}{[rad]}& &
\multicolumn{1}{c}{[mag]} & \multicolumn{1}{c}{[rad]}& & \multicolumn{1}{c}{[mag]} & \multicolumn{1}{c}{[rad]} \\
\noalign{\smallskip}
\hline
\noalign{\smallskip}
$f_1$ & & 17.21752 & & 0.1383 & 0.535 & & 0.136 &  0.50& &  0.138 &  0.53 & &  0.139 & 0.53\\
 & & $\pm$.00001 & & $\pm$.0003 & $\pm$.003 & & &  & & &  & & & \\
$f_2$ & & 22.30760 & & 0.0331 & 0.293 & & 0.033 &  0.27  & &    0.032 & 0.28 & &     0.034 &   0.29  \\
& & $\pm$.00003 & & $ \pm$.0003 & $\pm$.008 & &  &  & & &  & &  & \\
$2f_1$& &         & &  0.0316 & 4.968 & &     0.030 &     4.92 & &    0.031 &  4.95 & &     0.032 &  4.97\\
$f_1+f_2$& &      & &  0.0135 & 4.724 & &     0.013 &     4.65 & &    0.013 &  4.65 & &     0.014 &  4.75\\
$f_2-f_1$& &      & &  0.0096 & 2.875 & &     0.010 &     2.80 & &    0.007 &  3.02 & &     0.011 &  2.91\\
$3f_1$& &         & &  0.0084 & 3.205 & &     0.008 &     3.21 & &    0.008 &  3.22 & &     0.009 &  3.20\\
$2f_1+f_2$& &     & &  0.0055 & 2.874 & &     0.005 &     2.76 & &    0.005 &  2.76 & &     0.006 &  2.94\\
$2f_1-f_2$& &     & &  0.0046 & 1.350 & &     0.004 &     1.07 & &    0.004 &  1.26 & &     0.005 &  1.40\\
$4f_1$& &         & &  0.0030 & 1.524 & &            &           & &           &        & &     0.003 &  1.41\\
$3f_1+f_2$& &     & &  0.0026 & 1.223 & &            &           & &           &        & &     0.003 &  1.13\\
$3f_1-f_2$& &     & &  0.0023 & 5.621 & &            &           & &           &        & &     0.003 &  5.67\\
\noalign{\smallskip}
\multicolumn{3}{c}{Mean $V$ magnitude} &&\multicolumn{2}{c}{10.1980$\pm$.0002} &
& \multicolumn{2}{c}{}&& \multicolumn{2}{c}{} &
& \multicolumn{2}{c}{}\\
\multicolumn{3}{c}{Residual r.m.s. [mag]} &&\multicolumn{2}{c}{0.0111} & & \multicolumn{2}{c}{0.0078}&&
\multicolumn{2}{c}{0.0087} & & \multicolumn{2}{c}{0.0128}\\
\multicolumn{3}{c}{N}&&\multicolumn{2}{c}{4722} && \multicolumn{2}{c}{249}& &\multicolumn{2}{c}{1285} &&
\multicolumn{2}{c}{2421}\\
\noalign{\smallskip}
\hline
\end{tabular}
\label{sol}
\end{table*}

\begin{figure}
\resizebox{\hsize}{!}{\includegraphics{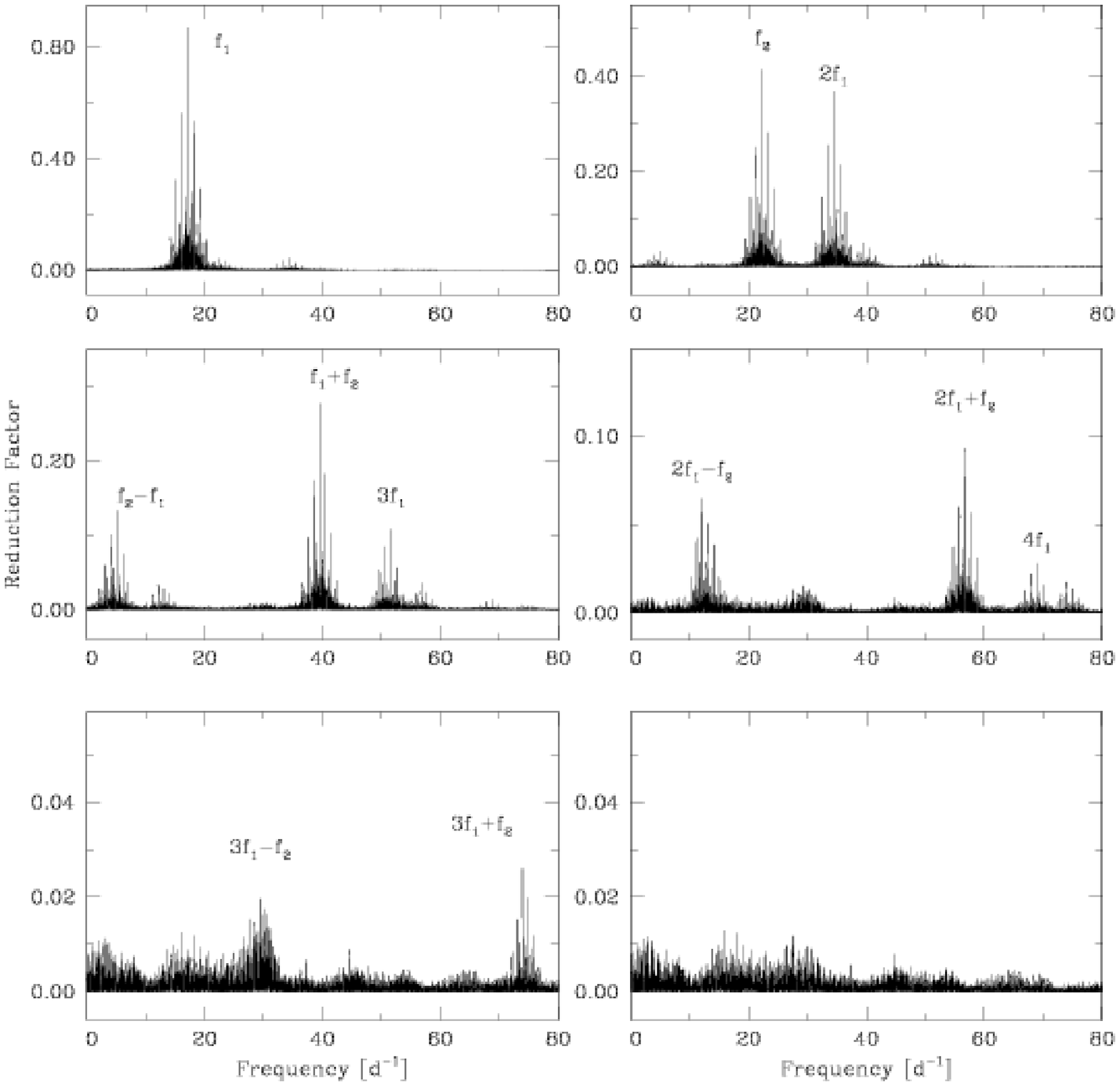}}
\caption[ ]{Power spectra of the \gsc data.}
\label{high}
\end{figure}
\begin{figure}
\resizebox{\hsize}{!}{\includegraphics{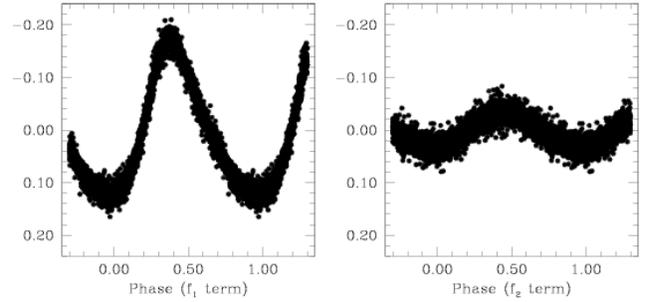}}
\caption[ ]{Light curves of the two independent terms found in the time series
of \gscp. The mean magnitude $V$=10.198 has been subtracted from the 4722 datapoints.}
\label{dm}
\end{figure}

Observations obtained
at two different longitudes (Europe and America) strongly reduced
the daily aliasing effect. To perform the various steps of the frequency analysis,
we used the least--squares power spectrum method (Vani\^cek 1971),
allowing us to detect one by one the constituents of the light curve. After each
detection, we refined the frequency values by applying the MTRAP code (Carpino et
al. 1987): the refined values  were introduced as known constituents in
the new search. Such a procedure is  particularly suitable for a multimode, high--amplitude
pulsation as it keeps the relationships between
the detected terms (i.e., 2$f_1$, $f_1+f_2$,~...) locked. Moreover, it does not require 
any data prewhitening as amplitudes and phases of the known constituents are recalculated
for each new trial frequency, always subtracting the exact amount of signal
for any detected term.

As a first step, we analyzed each subset separately.  
The frequency analysis of the 249 $uvby$ data obtained at S.~Pedro M\'{a}rtir 
(SPM subset) clearly detected eight terms: $f_1$=17.22~\cd, $f_2$=22.31~\cd,
2$f_1$, 3$f_1$, $f_1+f_2$, $f_2-f_1$, 2$f_1+f_2$, 2$f_1-f_2$. 
The term with the smallest amplitude is the 2$f_1-f_2$ one (0.0041 mag).
The power spectrum does not show any well--defined structure
after introducing these eight terms as known constituents, even though  
the peaks related to the  $f_1$ and $f_2$ coupling terms (3$f_1-f_2$, 3$f_1+f_2$,4$f_1$) 
are embedded among others,
without standing out clearly. The amplitude of the noise slightly increases toward low--frequencies
(0.0019 mag for $f<$5~\cd, 0.0016 mag for $f<$10~\cd), while it is quite small at high--frequencies
(0.0009 mag for 40$<f<$50~\cd). No difference was found in the frequency analysis
of different colours; the rms residuals are
0.0170, 0.0094, 0.0084 and 0.0078 in $u, v, b$ and $y$, respectively: the large 
scatter in $u$--light is due to some instrumental problems with that 
photomultiplier tube.

The analysis of the 1285 $V$ measurements carried out at Beersel Hills Observatory
(BHO subset) allowed us to detect the same terms as in the  SPM subset. 
The amplitude of the noise is 0.0008 mag in the 0$<f<$100~\cds interval, with
a small increase  at low frequencies (0.0015 mag for $f<$7\cd).
The solution with eight terms leaves an rms residual of
0.0087 mag. Other combination terms are visible in the residual spectrum, 
slightly higher than the noise
level. The magnitude differences between the comparison and checks star yield a 
standard deviation  of 0.010 mag; the values on individual nights range from 0.005 to 0.014 mag.

The data obtained at Athens University Observatory (AUO subset) are the most homogeneous ones. Eleven
terms have been detected: $f_1$, $f_2$,
2$f_1$, 3$f_1$, $f_1$+$f_2$, $f_2-f_1$, 2$f_1$+$f_2$, 2$f_1-f_2$ (i.e.,
those detected in the BH and SPM subsets) plus 3$f_1-f_2$, 3$f_1+f_2$, 4$f_1$
(i.e., the terms embedded in the noise in the BH and SPM  subsets). The solution 
with 11 terms leaves
an rms residual of 0.0128 mag: it is higher than that of previous subsets, but
the high number of datapoints (2421) allowed us to uncover terms with 
smaller amplitudes, by decreasing the amplitude  of the white noise (0.0007 mag in
the 0--100~d$^{-1}$ interval). The observed
scatter is similar to that observed for the check star GSC~0144-03140; 
the standard deviations range from 0.010 to 0.018 mag.

The three solutions of the $V$ or $y$ light curves are very coherent, detecting 
the same terms with very
similar amplitudes and phases (Table~\ref{sol}; a cosine series has been used).
Therefore, once the 
least--squares solutions for each subset were obtained, the mean magnitudes 
were aligned on the standard value $V$=10.198. 
At this stage, the other subsets (SETEC, OSN and MM) were added:
their samplings do not allow a reliable frequency analysis by themselves, but 
combined with  the previous ones they supply a better frequency
resolution. The alignments were made by applying the same procedure as above.
No systematic difference was found  between $V$ and $y$ data.
Hence, we have at our disposal a global dataset composed of 4722 measurements, which
constitutes a solid baseline for the careful search for small amplitude modes.

The analysis of the global dataset evidenced the large contribution of the $f_1$
term, i.e., 17.22~\cds (Fig.~\ref{high}, upper row, left panel). This periodicity
is characterized by a strong asymmetric light curve since the
harmonics 2$f_1$ (upper row, right panel), 3$f_1$ (middle row, left panel) and
4$f_1$ (middle row, right panel) were detected when going further in the analysis. 
The
second contribution in amplitude comes from the $f_2$=22.31~\cds term (upper row,
right panel): this term has a sine--shaped light curve, since no harmonic
was detected. The power spectra also show the contribution from the coupling
terms between the two independent frequencies: the $f_1+f_2$ and $f_2-f_1$ terms
(middle row, left panel), the $2f_1-f_2$ and $2f_1+f_2$ terms (middle row, right 
panel) and the $3f_1-f_2$ and $3f_1+f_2$ terms (lower row, left panel).
The last power spectrum (lower row, right panel) does not show any reliable 
residual peak. The mean noise level is around 0.52 mmag in the whole 0--80~\cds
interval, ranging from 0.39 mmag in the 50--80~\cds interval to
0.67 mmag in the 0--10~\cds one.

Table~\ref{sol} also summarizes the least--squares
solution of the global dataset.
We also inspected the residual light curves to see if there were systematic
deviations in some part of them, but we found nothing.
Formal errors of the two independent frequencies were  calculated 
following Montgomery \& O'Donoghue (1999).
Figure~\ref{dm} shows the light curve of each independent term. 
We stress the fact that no additional independent term with an amplitude
larger than 0.6 mmag was found: therefore, \gsc looks like
an authentic  pure  double--mode HADS star.

\section {The other double--mode HADS stars}

\begin{table*}
\caption{Frequencies of the $F$-- and 1O radial modes in double--mode HADS stars}
\begin{tabular}{l lll ll}
\hline
\hline
\noalign{\smallskip}
Star & F--mode & 1O mode & Ratio & & Reference \\
 & \multicolumn{1}{c}{[\cd]} & \multicolumn{1}{c}{[\cd]} &  & & \\
\noalign{\smallskip}
\hline
\noalign{\smallskip}
GSC 02583-00504      &  19.3347   & 25.0066   & 0.77318 &   & Wils et al. (2003)\\ 
SX Phe              &  18.19358 & 23.37940 & 0.77819 &   & Coates et al. (1980)\\  
GSC 00144-03031      &  17.21752 & 22.30760 & 0.77182 &   & This paper\\
OGLE BW2\_V142           &  15.14210  & 19.45340  & 0.77837 &   & This paper\\ 
BQ Ind              &  12.1951  & 15.7686  & 0.77338 &   & Sterken et al. (2003) \\
OGLE BW1\_V207           &  11.68214 & 15.09791 & 0.77376 &   & This paper and Moskalik \& Poretti (2002)\\
AE UMa              &  11.62560 & 15.03118 & 0.77343 &   & P\'{o}cs \& Szeidl (2001) \\   
Macho 116.24384.481 &  11.50557 & 14.88981 & 0.77271 &   & Alcock et al. (2000)\\  
Macho 115.22573.263 &  10.89872 & 14.11001 & 0.77241 &   & Alcock et al. (2000)\\ 
RV Ari              &  10.73788 & 13.89913 & 0.77258 &   & P\'{o}cs et al. (2002)\\  
GSC 06047-00749        &  10.08286  & 13.06908  & 0.77150 &   & This paper\\ 
Macho 114.19969.980 &\hskip 2.0truemm   9.68320 & 12.52957 & 0.77283 &   & Alcock et al. (2000)\\ 
Macho 119.19574.1169& \hskip 2.0truemm  9.35930 & 12.08875 & 0.77422 &   & Alcock et al. (2000); Moskalik \& Poretti (2002)\\ 
BP Peg              &\hskip 2.0truemm   9.1291 & 11.8329 & 0.77150 &   & Broglia (1959); Figer (1983) \\  
Macho 162.25343.874 &\hskip 2.0truemm   8.98622 & 11.64074 & 0.77196 &   & Alcock et al. (2000)\\ 
AI Vel              &\hskip 2.0truemm   8.96298 & 11.59958 & 0.77270 &   & Walraven et al. (1992)\\ 
Macho 128.21542.753 &\hskip 2.0truemm   8.32973 & 10.80603 & 0.77084 &   & Alcock et al. (2000)\\
Macho 119.19840.890 &\hskip 2.0truemm   7.96395 & 10.33167 & 0.77083 &   & Alcock et al. (2000)\\ 
V575 Lyr &\hskip 2.0truemm   6.86961 &  \hskip 2.0truemm 8.96848 & 0.76597 &   & Van Cauteren \& Wils (2001; GSC 02118-00297)\\
V703 Sco            &\hskip 2.0truemm   6.66838 &  \hskip 2.0truemm 8.67921 & 0.76832 &   & Oosterhoff (1966)\\
Brh V128            & \hskip 2.0truemm  6.519   & \hskip 2.0truemm 8.497   & 0.76721 &   & Bernhard et al. (2004)\\
HD 224852           &\hskip 2.0truemm   6.335592  & \hskip 2.0truemm 8.191974  & 0.77339 &   & This paper\\
GSC 03949-00811        &\hskip 2.0truemm   5.89097 & \hskip 2.0truemm 7.68763 & 0.76629 &   & This paper\\
GSC 04257-00471        &\hskip 2.0truemm   5.753783& \hskip 2.0truemm 7.514040& 0.76574 &   & This paper\\
VX Hya              &\hskip 2.0truemm   4.47648 & \hskip 2.0truemm 5.78972 & 0.77318 &   & Fitch (1966)\\
\noalign{\smallskip}
\hline
\label{hads}
\end{tabular}
\end{table*}

\begin{table*}
\caption{Parameters of the least--squares fits of new double--mode HADS stars.
For the stars in the lower part the 1O--mode has an amplitude larger than
the F--mode.}
\begin{tabular} {l c rrr c rrr c rrr  c rr }
\hline
\hline
\noalign{\smallskip}
& & \multicolumn{3}{c}{BW2\_V142}& & \multicolumn{3}{c}{BW1\_V207} & & \multicolumn{3}{c}{GSC 04257-00471}\\
\cline{3-5} \cline {7-9} \cline {11-13}\\
\multicolumn{1}{c}{Term} & & \multicolumn{1}{c}{Freq.} & \multicolumn{1}{c}{Ampl.} & \multicolumn{1}{c}{Phase} & &
\multicolumn{1}{c}{Freq.} & \multicolumn{1}{c}{Ampl.} & \multicolumn{1}{c}{Phase} & &
\multicolumn{1}{c}{Freq.} & \multicolumn{1}{c}{Ampl.} & \multicolumn{1}{c}{Phase}\\
\multicolumn{1}{c}{} & & \multicolumn{1}{c}{[\cd]} &  \multicolumn{1}{c}{[mag]} & \multicolumn{1}{c}{[rad]} & &
\multicolumn{1}{c}{[\cd]} &  \multicolumn{1}{c}{[mag]} & \multicolumn{1}{c}{[rad]} & &
\multicolumn{1}{c}{[\cd]} &  \multicolumn{1}{c}{[mag]} & \multicolumn{1}{c}{[rad]}\\
\noalign{\smallskip}
\hline
\noalign{\smallskip}
$f_1$ & & 15.14210 &      0.145 &     4.84& & 11.68214    &     0.138 &     6.13 & & 
           5.75378 &      0.082 &     1.82 \\
& & $   \pm$.00002 &  $\pm$.006 & $\pm$.04& & $\pm$.00001 & $\pm$.004 & $\pm$.03& &
       $\pm$.00008 &  $\pm$.003 & $\pm$.04\\
$f_2$ & & 19.45340 &  0.070 & 2.28     & & 15.09791 & 0.022 & 0.04 & &
           7.51404 &      0.033 &     1.41\\
& & $\pm$.00005    &$\pm$.006 & $\pm$.08&& $\pm$.00008 & $\pm$.004 & $\pm$.17& &
    $\pm$.00024    &$\pm$.003 & $\pm$.10 \\
$2f_1$& &          &     0.047 & 0.80  & &         & 0.049 & 3.88 & &  & 0.027 & 1.55\\
$f_1+f_2$& &       &   0.029 & 4.99    & &         &       &      & &  & 0.027 & 1.02\\
$f_2-f_1$& &       &   0.035 & 0.24    & &         &       &      & &  &  & \\ 
\noalign{\smallskip}
\multicolumn{2}{l}{Mean magnitude} & \multicolumn{3}{c}{$I$=17.842$\pm$.004} & &  
 \multicolumn{3}{c}{$I$=17.852$\pm$.003} & & \multicolumn{3}{c}{$R_{\rm NSVS}$=11.108$\pm$.002}\\
\multicolumn{2}{l}{Residual r.m.s.} &  \multicolumn{3}{c}{0.042 mag}  & &
\multicolumn{3}{c}{0.034 mag} & & \multicolumn{3}{c}{0.040 mag}\\
\multicolumn{2}{l}{$T_0$}  &\multicolumn{3}{c}{HJD 2449234.4035} & &
\multicolumn{3}{c}{HJD 2449249.8878} & & \multicolumn{3}{c}{MJD 2451428.0670}\\
\multicolumn{2}{l}{N} & \multicolumn{3}{c}{120} & &
\multicolumn{3}{c}{171} & & \multicolumn{3}{c}{309}\\
\noalign{\smallskip}
\hline
\noalign{\smallskip}
& & \multicolumn{3}{c}{GSC 06047-00749}& & \multicolumn{3}{c}{HD 224852} & & \multicolumn{3}{c}{GSC 03949-00811}\\
\cline{3-5} \cline {7-9} \cline {11-13}\\
\multicolumn{1}{c}{Term} & & \multicolumn{1}{c}{Freq.} & \multicolumn{1}{c}{Ampl.} & \multicolumn{1}{c}{Phase} & &
\multicolumn{1}{c}{Freq.} & \multicolumn{1}{c}{Ampl.} & \multicolumn{1}{c}{Phase} & &
\multicolumn{1}{c}{Freq.} & \multicolumn{1}{c}{Ampl.} & \multicolumn{1}{c}{Phase}\\
\multicolumn{1}{c}{} & & \multicolumn{1}{c}{[\cd]} &  \multicolumn{1}{c}{[mag]} & \multicolumn{1}{c}{[rad]} & &
\multicolumn{1}{c}{[\cd]} &  \multicolumn{1}{c}{[mag]} & \multicolumn{1}{c}{[rad]} & &
\multicolumn{1}{c}{[\cd]} &  \multicolumn{1}{c}{[mag]} & \multicolumn{1}{c}{[rad]}\\
\noalign{\smallskip}
\hline
\noalign{\smallskip}
$f_1$ & & 13.06908 &      0.101 &   2.86 & & 8.191974    &     0.116 &  0.15 & & 7.68763 &      0.094 &     1.84 \\
& & $   \pm$.00001 &  $\pm$.002 & $\pm$.02& & $\pm$.000006 & $\pm$.001 & $\pm$.01& & $\pm$.00007 &  $\pm$.004 & $\pm$.04\\
$f_2$ & & 10.08286 &  0.082 & 2.18     & & 6.335592 & 0.086 & 5.88 & & 5.89097 &      0.074 &     4.58\\
& & $\pm$.00001    &$\pm$.002 & $\pm$.03& & $\pm$.000005 & $\pm$.001 & $\pm$.01& & $\pm$.00009    &$\pm$.003 & $\pm$.05 \\
$2f_1$& &          &     0.019 & 3.71  & &         & 0.026 & 4.94 & &  & \\
$f_1+f_2$& &       &   0.024 & 3.03    & &         & 0.037 & 3.93 & &  & 0.032 & 4.69\\
$f_1-f_2$& &       &         &         & &         & 0.022 & 3.53 & &  & 0.022 & 0.51\\
$2f_2$   & &       &         &         & &         & 0.017 & 3.16 & &  &       &     \\
$3f_1$   & &       &         &         & &         & 0.008 & 2.48 & &  &       &     \\
$2f_1+f_2$&&       &         &         &&          & 0.009 & 0.75 & &  &       &     \\
$2f_1-f_2$&&       &         &         & &         & 0.009 & 1.45 & & &        &     \\
$f_1+2f_2$&&       &         &         & &         & 0.007 & 1.54 & & &        &     \\
\noalign{\smallskip}
\multicolumn{2}{l}{Mean magnitude}  &\multicolumn{3}{c}{$V$=12.012$\pm$.001} & &
 \multicolumn{3}{c}{$V$=10.131$\pm$.001} & & \multicolumn{3}{c}{$R_{\rm NSVS}$=11.128$\pm$.003}\\
\multicolumn{2}{l}{Residual r.m.s.} &  \multicolumn{3}{c}{0.030 mag}  & &
\multicolumn{3}{c}{0.023 mag} & & \multicolumn{3}{c}{0.034 mag}\\
\multicolumn{2}{l}{$T_0$}  &\multicolumn{3}{c}{HJD 2449236.3120} & &
\multicolumn{3}{c}{HJD 2453221.7986} & & \multicolumn{3}{c}{MJD 2451247.6966}\\
\multicolumn{2}{l}{N}  &\multicolumn{3}{c}{670} & &
\multicolumn{3}{c}{1873} & & \multicolumn{3}{c}{202}\\
\noalign{\smallskip}
\hline
\hline
\end{tabular}
\label{bw}
\end{table*}

The class of double--mode HADS stars has been enriched with various 
new entries provided by large--scale projects. 
In addition to \gscp, we discovered new double--mode pulsators when
performing systematic analyses of some public databases:
OGLE (Optical Gravitational Lensing Experiment; 
Szyma\'{n}ski 2005 and Udalski et al., 1997), NSVS
(Northern Sky Variability Survey; Wo\'{z}niak et al. 2004),
ASAS (All Sky Automated Survey; Pojmanski 2002 and 2003)
In some cases, we planned new photometric observations.  
The enlarged sample allows us to define
better the characteristics of the double--mode HADS stars.
As shown by the analysis of \gscp, there are double--mode radial pulsators
which do not show any hint of nonradial terms, even when the amplitude
threshold is very small. On the other hand, Poretti (2003) showed
how the radial double--mode pulsation can be also paired to a nonradial one.

Table~\ref{hads} lists the stars where the $F$ and 1O modes have been
detected. The formal errors on the frequency values are usually better than
1$\cdot10^{-4}$~\cd, which in turn implies an error on the ratio values better than
1$\cdot10^{-5}$. The new variables are described below.

\subsection{BW2\_V142 and BW1\_V207} 
Moskalik \& Poretti (2002) reported on the double--mode nature
of some stars contained in the OGLE--I database. We note that
BW1\_V109$\equiv$Macho 119.19574.1169 (Alcock et al. 2000).
The first least--squares solutions of the BW2\_V142 and BW1\_V207 light
curves are given in Table~\ref{bw}. 
Poretti (2000) briefly reported on BW2\_V142
just showing the light curves on the two terms (note that
there is a misprint in Fig.~1 about the value of the $f_1$ term). 
\gsc and
BW2\_V142 fill the gap between shortest periods  (namely those
of SX Phe and GSC 02583--00504) and the
group of HADS stars which pulsate with periods around 0.10~d. 

\subsection{GSC 03949-00811} The star is catalogued as NSVS~3234596 by Wo\'{z}niak et al. (2004).
The frequency analysis of the NSVS data  allowed us to  detect the $f_1$=7.687632~\cd,
$f_2$=5.890971~\cd, $f_1+f_2$ and $f_1-f_2$ terms. The amplitude of the
1O--mode is larger than that of the $F$--mode. The scatter 
is quite high (0.034 mag) and it was not possible to evidence other coupling
or additional terms.
The least--squares solution of the NSVS light curve is reported in Tab.~\ref{bw}.

\subsection{GSC 06047-00749} 
 Three subsets are available for this star: ASAS
(ASAS3 094303-1707.3; Pojmanski 2002 and 2003), 
NSVS (NSVS~15731136; Wo\'{z}niak et al. 2004) and a new
one obtained by NB (0.20--m telescope equipped with a SBIG ST7e CCD and
a $V$ filter, 402 measurements on four nights in June, 2004).
The terms $f_1$=13.0691~\cd, $f_2$=10.0829~\cds and $f_1+f_2$ were
detected in all the subsets. The least--squares solutions supplied
a smaller amplitude for the NSVS subset, since these data are in unfiltered
light.
The solution reported in Table~\ref{bw} was obtained by combining
the ASAS and the NB subsets. 

\subsection{HD 224852} 
The variability of HD~224852$\equiv$NSV 14800 was discovered
by Strohmeier \& Ott (1967) and, independently, by ASAS
(ASAS3 000116-6037.0; Pojmanski 2002 and 2003). In
the framework of our project, the star
was observed by NB (1671 measurements on nine nights from October to December, 2004). 
The $f_1$=8.1920~\cd, $f_2$=6.3356~\cd, 2$f_1$, $f_1+f_2$, $f_1-f_2$ terms 
were detected in the ASAS subset.
The observations carried out by NB are more dense and also the 2$f_2$, 2$f_1+f_2$
terms were detected. In spite
of the small differences in the amplitudes between the two subsets, the
frequency analysis of the combined dataset evidenced  
other coupling terms: 3$f_1$, 2$f_1-f_2$, $f_1+2f_1$. 
The search for new terms was stopped when the noise at the lowest frequencies
became predominant.
The 1O--mode has an amplitude larger than the
$F$--one and the light curves of both periods have a non--sinusoidal shape. 
Table~\ref{bw} lists the least--squares solution of the combined dataset;
the mean magnitude has been aligned to that of the ASAS subset. The solution 
with 10 terms leaves an rms residual of 0.023 mag; the amplitude of the noise
is 3.3 mmag for $f<$5~\cd, 1.6 mmag for 10$<f<$50~\cd. Also in this case there is 
no hint of the excitation of additional modes.
\subsection{GSC 04257-00471}
The star is catalogued as NSVS~3324715$\equiv$NSVS~3361415 by Wo\'{z}niak et al. (2004). 
We could detect the $f_1$=5.7538~\cd,
$f_2$=7.5140~\cd, 2$f_1$ and  $f_1-f_2$ terms (Table~\ref{bw}). 
The amplitude of the 1O--mode is quite small, but the NSVS data are in unfiltered light and
therefore damping is expected with respect to $V$--light. We note that GSC~03949-00811
and GSC~04257-00471  fill an important gap toward long periods. 


\begin{figure}
\resizebox{\hsize}{!}{\includegraphics{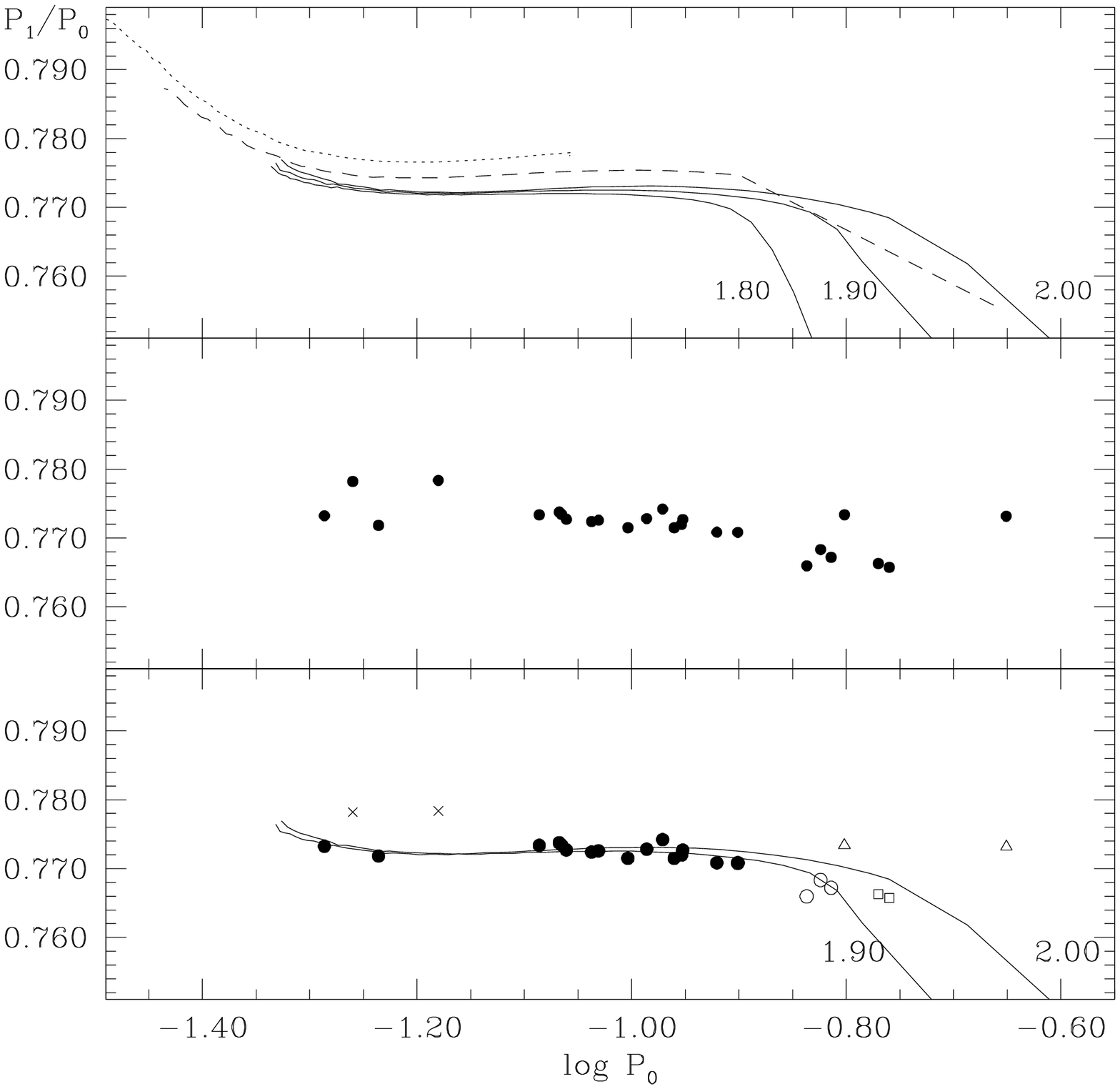}}
\caption[ ]{Petersen's diagram for double--mode HADS stars. The {\it upper
panel} shows the sequences calculated from our models.
Solid lines: [Fe/H]=0.00 dex and three different masses (from 1.80 to 2.00~M$_{\sun}$);
dashed line: [Fe/H]=--0.50 dex and M=1.80~M$_{\sun}$;
dotted line: [Fe/H]=--1.00 dex and M=1.80~M$_{\sun}$.
The {\it middle panel} shows the observed period ratios. The {\it lower panel}
illustrates  the
matching between theory and observations: the solid lines are 
the theoretical sequences calculated for [Fe/H]=0.00 dex, M=1.90~M$_{\sun}$ and
[Fe/H]=0.00 dex, M=2.00~M$_{\sun}$.
The crosses indicate low--metallicity (Pop.~II) stars; the open symbols (circles, squares and
triangles) indicate Pop.~I stars with different masses (1.90, 2.00 and $>$2.00~M$_{\sun}$,
respectively). The filled circles indicate the Pop.~I stars in which the differences in the 
physical parameters, if any, generate similar period ratio values.}
\label{pete}
\end{figure}
\begin{figure}
\resizebox{\hsize}{!}{\includegraphics{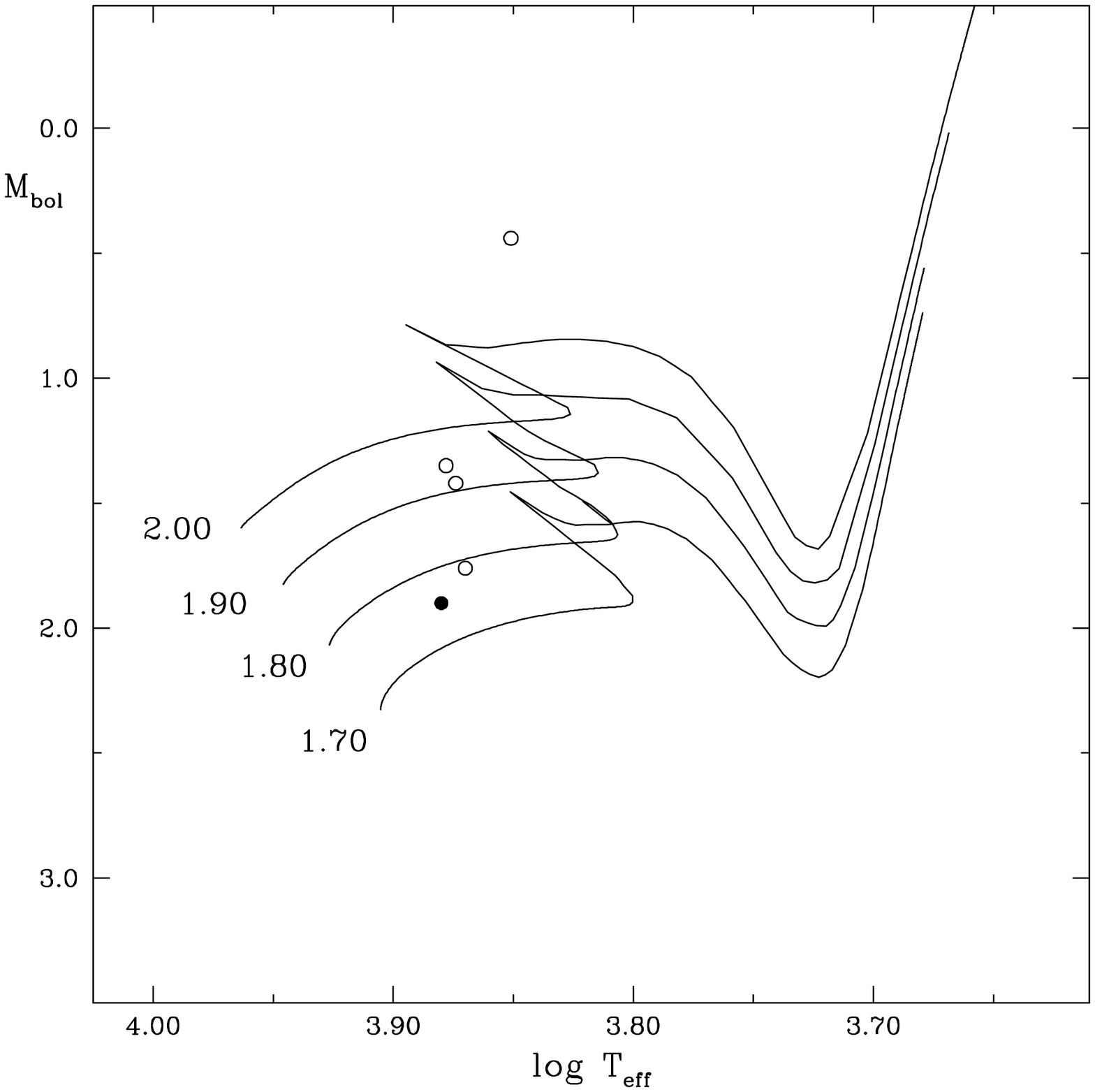}}
\caption[ ]{Evolutionary tracks calculated for [Fe/H]=0.00 dex,  $d_{\rm over}=0.2$ and
four different values for the mass (from 1.70 to 2.00 M$_{\sun}$).
The filled circle indicates the position of 
\gscp.  The open circles, from bottom to top, indicate RV Ari, AI Vel, BP Peg and
VX Hya.}
\label{tracks}
\end{figure}

\section{Modelling the double--mode HADS stars }
The period ratios reported in Table~\ref{hads} are plotted in the middle panel of Fig.~\ref{pete}.
Most of the stars are grouped around $\log~P_0=-1.0$ and $P_1/P_0$=0.772. 
It appears that the period ratio decreases toward  the longest periods. However,
some deviating points are observed at both extrema. In particular, the decrease
in the period ratio values around  $\log~P_0=-0.8$ appears much more rapid than
the one  calculated by Petersen \& Christensen-Dalsgaard (1996; their Fig.~4) using
the star sample available at that time. Therefore, we calculated a new set of 
theoretical models.
\subsection{The theoretical sequences in the Petersen diagram}
Stellar models were computed with the evolutionary code CESAM
(Morel 1997). The numerical precision as well as the model mesh grid
(2000 mesh points approximately) were optimized to compute oscillations.
The equation of state CEFF (Christensen-Dalsgaard \& Daeppen 1992) was 
used, also including the Coulombian
correction to the classical EFF (Eggleton et al. 1973).
A weak electronic screening
in these reactions was assumed (Clayton 1968), accordingly to 
the evolutionary stages considered in this work.
Opacity tables were taken from the OPAL package (Iglesias \& Rogers 1996), 
complemented at
low temperatures ($T\leq10^4\,$K) by the tables provided by
Alexander \& Ferguson (1994). For the atmosphere reconstruction, the
Eddington~$T(\tau)$ law (grey approximation) was considered.
                                                                                                                    
The transformation from heavy element abundances with respect to
Hydrogen [M/H] into concentration in mass $Z$ assumes an
enrichment ratio of $\Delta Y/\Delta Z=2$ and $Y_{\mathrm{pr}}=0.235$ and
$Z_{\mathrm{pr}}=0$ for Helium and heavy element primordial concentrations,
respectively.
Convective transport was described by the classical Mixing Length theory,
with efficiency and core overshooting parameters set to
$\alpha_{ML}=l_{m}/\mathrm{H}_p=1.8$ and $d_{\rm over}=l_{\rm over}/\mathrm{H}_p=0.2$,
respectively. The latter parameter is prescribed by Schaller et al. (1992) for
intermediate mass stars. $\mathrm{H}_p$ corresponds to the local pressure
scale-height, while $l_{m}$ and $l_{\rm over}$ represent respectively the mixing length
and the inertial penetration distance of convective elements.
                                                                                                                    
Theoretical oscillation spectra were computed from the equilibrium models
described above. For this purpose the oscillation code
\emph{Filou} (Su{\' a}rez 2002) was used.
The upper panel of Fig.~\ref{pete}  shows various sequences
calculated from our models.
The effects of changing the mass  keeping [Fe/H]=0.00 dex are shown
by the three solid lines. The other two lines show the effect
of decreasing metallicity keeping M=1.80~M$_{\sun}$: from [Fe/H]=--0.50 dex
(dashed line) to [Fe/H]=--1.00 dex (dotted line).

\subsection{The evolutionary tracks}
The physical parameters of HADS stars are summarized   
by McNamara (2000). $uvby\beta$ photometry of \gsc 
($V$=10.198,  $b-y$=+0.138, $m_1$=0.194, $c_1$=0.849, $\beta$=2.805;
SPM and OSN observations described above) yields the following parameters:
E$_{b-y}$=+0.005, M$_V$=+2.00, T$_{\rm eff}$=7590~K, $\log~g$=3.98, 
[Fe/H]=--0.05 dex (TEMPLOGG method; Rogers 1995, Kupka \& Bruntt 2001).

The connection between double--mode HADS stars is
confirmed when considering the evolutionary tracks.
The tracks have been calculated for [Fe/H]=0.00 dex, a value
which can be applied to \gscp, RV Ari ([Fe/H]=0.00 dex), BP Peg ($-0.02$),
AI Vel ($-0.15$) and VX Hya (0.05).
Figure~\ref{tracks} shows the positions in the $\log~$T$_{\rm eff}$--M$_{\rm bol}$ 
plane for these stars.
The position of \gsc suggests M=1.75~M$_{\sun}$; such a position is   
similar to that of RV Ari. 
Note how the points in Fig.~\ref{tracks} describe a narrow instability
strip for double--mode HADS stars.

\section{Interpreting the Petersen diagram}
The decreasing trends in the theoretical sequences are characterized
by  a long standstill between $\log~P_0=-1.30$ and 
$\log~P_0=-0.90$ (Fig.~\ref{pete}, upper panel).
In this interval the $P_1/P_0$ values are more
sensitive to the changes in metallicity than to changes in mass. On
the other hand, small differences in mass produce strong differences
in the sequences at the longest periods. We can compare these predictions
with the Petersen diagram as determined from the stars listed in Table~\ref{hads}.

\subsection{Stars with $P_0$=0.10~d}

The period ratios  cluster  around 0.772 for $\log~P_0=-1.0$.
This point in the Petersen diagram is a well--defined one on the
basis of the enlarged sample: the theoretical models have to match it.
We calculated a large variety of models, but a satisfactory match is provided only
for [Fe/H]=0.00 dex and 1.70$<$M$<$2.00~M$_{\sun}$. These constraints
are the same obtained from the positions of double--mode HADS stars in the
$\log~$T$_{\rm eff}$--$M_{\rm bol}$ plane (Fig.~\ref{tracks}). We note that the changes
in metallicity  shift the theoretical sequences toward higher period ratio values
(Fig.~\ref{pete}, upper panel). Therefore, the small dispersion of the points related
to stars with $P_0$=0.10~d suggests a very similar metallicity (Fig.~\ref{pete}, middle panel). 
The period ratio value corresponding to AE UMa (Table~\ref{hads}) is quite normal
(0.7738), even though AE UMa is a 
slightly metal deficient star ([Fe/H]=$-0.50$~dex).  However, it is also less
massive than other stars (M=1.49~M$_{\sun}$; McNamara 2000): the two effects balance
each other and generate the normal period ratio value.

\subsection{The shortest periods}
The extreme values observed at the shortest periods refer to \gscp, GSC 02583--00504,
SX Phe and BW2\_V142. 
If we admit that stars with $\log~P_0=-1.0$ and
$P_1/P_0$=0.772 constitute a homogeneous sample, \gsc and GSC 02583--00504 can be
considered as the natural extension toward the shortest period of this sample.
Indeed, the sequences  with [Fe/H]=0.00 dex and  masses in the 1.70--2.00~M$_{\sun}$ range
match very well also these points (Fig.~\ref{pete}, solid lines in the upper
panel).
The high period--ratio values shown by SX Phe and BW2\_V142 (crosses
in the lower panel of Fig.~\ref{pete}) 
can be explained by a lower metallicity (Fig.~\ref{pete}, dotted line in the upper
panel), which is 
a well--established fact for SX Phe ([Fe/H]=--1.40 dex, McNamara 2000), the prototype of  Pop.~II
short--period pulsating stars.  The same argument holds for BW2\_V142,
as this star is located in the galactic bulge. 
We can infer that the spread of the period ratio values 
at the shortest periods can be explained by  different metallicities.

\subsection{The longest periods}
V575 Lyr, V703 Sco, BrhV128, GSC~04257-00471 and GSC~03949-00811 are the stars defining the
rapid decrease at $\log~P_0>-0.90$.
This decrease is in agreement 
with the  theoretical sequences and hence these
stars are naturally linked to the ones at $P_0$=0.10~d.
However, the smaller the mass, the  more rapid  the decrease is
(Fig.~\ref{pete}, solid lines in the upper panel).
Therefore, only the M=1.90~M$_{\sun}$ models are able to match the
points related to V575 Lyr, V703 Sco and BrhV128
(Fig.~\ref{pete}, open circles in the lower panel) and
only the M=2.00~M$_{\sun}$ models are able to match the
points related to GSC~04257-00471 and GSC 03949-00811
(Fig.~\ref{pete}, open squares in the lower panel).

Two long--period stars show period ratios  higher than 0.770,
i.e., HD~224852 and VX Hya (triangles in the lower 
panel of Fig.~\ref{pete}). Such values are admitted by models
with M$>2.00$~M$_{\sun}$ only. Indeed, 
VX Hya is a very luminous (see Fig.~\ref{tracks}) and massive
(more than 2.30~M$_{\sun}$, McNamara 2000)  star. It is for sure a
Pop.~I star ([Fe/H]=+0.05 dex) and therefore we can rule out a connection
with the Pop.~II stars SX Phe and BW2\_V142.

Therefore the spread in the period ratio values observed at the longest
periods can be explained by differences in the masses. These differences
in masses cannot be distinguished at $\log~P_0<-0.90$, where
the sequences are almost coinciding.
We note that the rapid decrease in the $P_1/P_0$ values for M$<1.80$~M$_{\sun}$
could explain some unusual ratio (as for example Macho~104.20389.1202,  
$\log~P_0=-0.80$, $P_1/P_0$=0.751; Alcock et al. 2000), 
without invoking the excitation of nonradial modes (see Fig.~5 in Poretti 2003). 
\section{Conclusion}
The new observations of \gsc confirm the existence of pure radial double--mode
pulsators. Thanks to large--scale projects it has been possible to discover
several new double--mode HADS stars and to discuss the properties of the class.
Detailed models have been calculated and compared with the
observed period ratios and the physical parameters determined on the basis of  $uvby\beta$ 
photometry.
Our models suggest that double--mode HADS stars constitute a very homogeneous
class of variable stars.  The Petersen diagram for these
stars can be reconstructed using models having [Fe/H]=0.00 dex
and M=1.70--2.00M$_{\sun}$. Different masses generate similar period
ratios  for $\log~P<-$0.90, but theoretical period ratios predict 
differences for  $\log~P>-$0.90. More massive
stars show higher values: this spread is actually
observed in the Petersen diagram. The sequences  with 1.90$<$M$<$2.00~M$_{\sun}$ match
most of the observed period ratios in a very satisfactory way. 
On the other hand, 
short--period Pop.~II stars show slightly higher period ratios owing to their
different metallicity and they clearly deviate from the Pop.~I sequence.

\begin{acknowledgements}
The authors wish to thank the STARE team for the attribution of observing time
to the COROT project. 
J.~Vialle checked the English form of the manuscript.
This publication makes use of data partially taken from the Northern Sky Variability Survey 
created jointly by the Los Alamos National Laboratory and
University of Michigan. The NSVS was funded by the Department of Energy, the 
National Aeronautics and Space Administration, and the National Science Foundation.
Part of the collected data were acquired with equipment purchased thanks to
a research fund financed by the Belgian National Lottery (1999).
PGN, KDG and VNM acknowledge financial support from the Special Account for Research
Grants No 70/3/7187 (HRAKLEITOS) and 70/3/7382 (PYTHAGORAS) of the
National and Kapodistrian University of Athens, Greece.
RA and JAB acknowledge financial support from grants
AyA2001-1571 and ESP2001-4529-PE of the Spanish National Research plan.
JCS acknowledges the financial support from the  Spanish  Plan of
Astronomy and Astrophysics, under
project $AYA2003-04651$, from the Spanish ``Consejer\'{\i}a de
Innovaci\'on, Ciencia y Empresa",  from the
 ``Junta de Andaluc\'{\i}a" local government, and from the Spanish Plan
Nacional del Espacio under project ESP2004-03855-C03-01.
EP acknowledges financial support from PRIN 2003 ``Continuit\`a e discontinuit\`a
nella formazione della nostra Galassia".
\end{acknowledgements}

\end{document}